# An Overview of High-Altitude Balloon Experiments at the Indian Institute of Astrophysics

Margarita Safonova,* Akshata Nayak,† A. G. Sreejith, Joice Mathew, Mayuresh Sarpotdar,
S. Ambily, K. Nirmal, Sameer Talnikar, Shripathy Hadigal,‡ Ajin Prakash and Jayant Murthy

## Abstract

We have initiated the High-Altitude Ballooning programme at Indian Institute of Astrophysics, Bangalore, in the year 2011 with the primary purpose of developing and flying low-cost scientific payloads on a balloon-borne platform. The main aim is the observations of extended nearby objects (e.g. comets) and of diffuse sources (e.g. zodiacal light or airglow) with wide field of view (FOV) UV instruments from near space (20 to 30 km). A brief summary and the results of the tethered flights carried out at IIA CREST campus are given in Ref. 1. Here we present an overview of the nine free-flying balloon experiments conducted from March 2013 to November 2014. We describe the launch procedures, payloads, methods of tracking and recovery. Since we fall in the light balloon category — payload weight is limited to less than 6 kg — we use the 3-D printer to fabricate lightweight boxes and structures for our experiments. We discuss the flight and scientific data obtained from different launches and the development of the in-house lightweight sensors and controllers, as well as a fully-fledged 2-axis pointing and stabilization system, for the flights.

*For correspondence (E-mail: rita@iiap.res.in)
†Akshata Nayak is a Ph.D. student at the Jain University, Bangalore 562 112, India,
‡Sameer Talnikar and Shripathy Hadigal at the time of writing were internship students at the Indian Institute of Astrophysics, Bangalore 560 034, India.

## 1 Introduction

Near space is the region of Earth's atmosphere that lies between 20 to 100 km above sea level, encompassing the stratosphere, mesosphere, and the lower thermosphere. There is a surge of interest in flying balloons to the edge of near space in recent years, largely for recreational and educational purposes due to the major reduction in costs of all the involved components, from inexpensive latex balloons to easily available lightweight, compact, and simple in operation micro-electromechanical (MEM) devices. However, the same availability allows also to apply this to the scientific objectives, to do serious science at low cost. Though space observatories provide accurate observations of remote or faint space sources, they are also prohibitively expensive and only affordable to large governmental agencies. Small telescopes/cameras onboard balloons or sounding rockets are attractive because they are much cheaper and yet can yield substantial scientific output; the first UV spectrum of a quasar was obtained during a short rocket flight (Davidsen et al., 1977). We have initiated (Sreejith et al., 2012) the High-Altitude Ballooning (HAB) programme at Indian Institute of Astrophysics (IIA), Bangalore, in the year 2011 with the primary purpose of developing and flying low-cost scientific payloads on a balloon-borne platform; an endeavor to enable carrying out scientific space experiments and observations at costs accessible to university departments.

Our science goals include studies of the phenomena occurring in the upper atmosphere in the ultraviolet (UV) range, of airglow and zodiacal light, and spectroscopic UV observations of extended astronomical objects such as, for example, comets, using the spectrograph in the near-UV window from 200 to 400 nm; the range that has been yet largely unexplored by the balloon (or the UV) community. This window includes the lines from several key players in atmospheric chemistry such as $S0_2$, $O_3$, BrO, HCHO and allows to observe a strong OH signal (308 nm) in astronomical sources (also $C_2$, CS, $CO_2^+$). We are also planning performing high-altitude astrobiological experiments, such as for example survival of microbes in upper atmosphere (David, 2013), or collection of stratospheric samples to study the air/dust composition, especially in view of the recent claims of detection of extra-terrestrial cells in stratosphere (Narlikar et al., 2003). Despite the importance of this topic to astrobiology (stratospheric values of gas pressure, temperature, humidity and radiation are very similar to the surface of Mars), stratospheric microbial diversity/survival remains largely unexplored (Smith et al., 2010), probably due to significant difficulties in the access and in ensuring the absence of ground–mid-atmospheric contaminations.

Alongside our science objectives, we also intent on developing our own experimental and scientific equipment for balloon-borne and eventual space flights. A key requirement for HAB observations is the accurate and stable pointing platform, where the data can be stored for later retrieval or transmission. The challenge is in doing this for an experiment where weight and power are strictly limited. We have designed and manufactured a 2-axis stabilization system (Nirmal et al., 2015) to correct for payload random motions, where the crucial step is determination of the pointing direction (Sreejith et al., 2014). For additional accuracy, we are developing a low-cost star tracker, where the attitude sensor is used for the initial coarse pointing (Sarpotdar et al., 2014). Development of the image-intensified UV detector (Ambily et al., 2015) on telescopic system for spectroscopic (Sreejith et al., 2015) and imaging applications to fly onboard the balloon payload is also underway.

Approvals from different airport and airforce agencies are necessary to perform free-flying balloon experiments. These permissions have to be requested at least 8 to 12 months before the beginning of the program. We have applied for these permissions in 2012, and have conducted our first free-flying balloon experiment on March 3, 2013. In 2013–2014, we have carried out a total of nine launches from CREST campus of IIA, Hoskote region, Karnataka. There have been failures in recovery, working of the electronic systems and even loose connections in wires during flights. But we have learned from our experiences and put together some successful flights. During this time, we continued a parallel program of developing the in-house low-cost scientific and technological equipment for our HAB programme. The following sections give a detailed overview of these 9 launches.

## 2 Essentials for the balloon flights

### 2.1 Permissions necessary to carry out the launches

Since Bangalore is a hub of AirForce bases along with airports and flying training schools, we had to take prior permissions from the Director General of Civil Aviation (DGCA), Ministry of Defence (MoD), and Airports Authority of India (AAI), Delhi. In addition, we require local permissions from Hindustan Aeronautics Limited (HAL), Bangalore International Airport Limited (BIAL), Chennai airport as a southern region headquarters, Jakkur flying school, Yelahanka AirForce Base, AirForce station at Chimney hills, Bangalore. It is a regulation that we inform the abovementioned offices about the balloon launch two weeks in advance, specifying the payload details. Launches are only allowed after obtaining all the required No-Objection Certificates (NOCs), and we can carry out the flights only on a certain local non-flying days, currently on Sundays.

### 2.2 Balloons

The overall structure of the HAB system is presented in Fig. 1: balloon(s) on the top, parachute(s) between the balloon and the payload, and the payload(s) at the bottom.

Many types of balloons are used for high-altitude balloon experiments depending on the size of the pay-load, scientific objectives and the scale of the programme. Latex balloons, also known as sounding or weather balloons, are designed to reach 35–40 km and burst, after which a parachute is deployed to safely carry the payload back to Earth. They are low-cost, easily available and, since they usually do not exceed volumes of 3 $m^3$ on the ground, are easy and inexpensive to launch. There are also thin plastic balloons, called zero-pressure balloons, which can float at high altitudes for days, reach $\sim$ 50 km altitudes and carry payloads of several tons. They are also very expensive, require special manufacturing and specially designed launch towers. They are usually employed by large-scale institutions, such as NASA's Balloon Program, National Scientific Balloon Facility, USA (Smith, 2002), Tata Institute of Fundamental Research (TIFR) National Balloon Facility, Hyderabad, India (Vasudevan et al., 2012) or the Japanese balloon base at TARF, Japan (Fuke et al., 2010). Since it is our aim to develop the low-cost balloon programme affordable to the universities and colleges in India, we decided to use latex balloons as they are cheap, small and readily available. We initially used 3-kg balloons from Ningbo Yunhan Electronics Co. Ltd. (China), but eventually switched off to 1.2 and 2-kg balloons from Pawan Exports, Pune, India, as a cheaper and easily obtainable alternative.

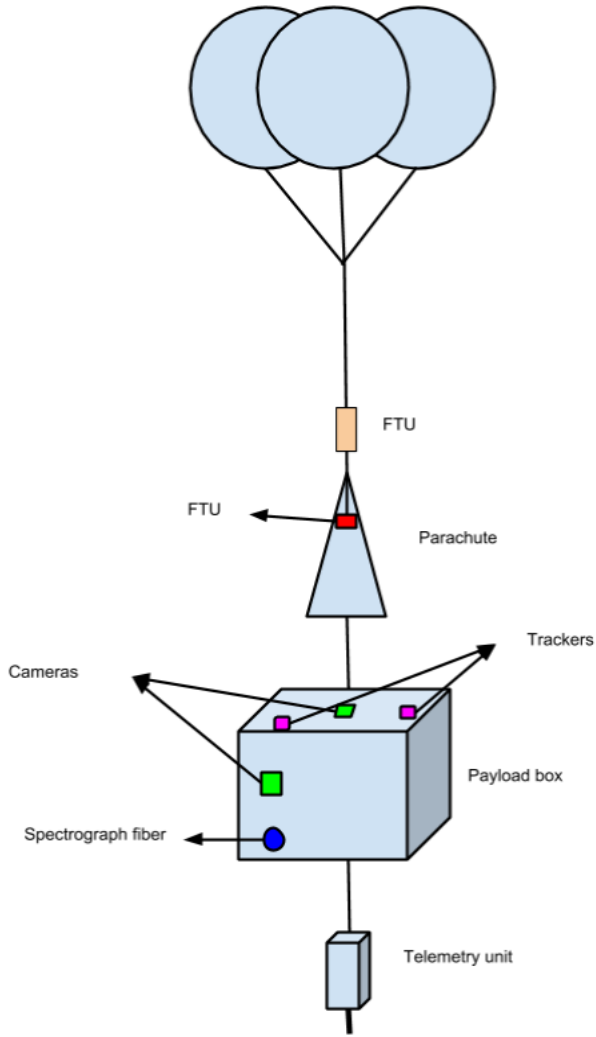

Figure 1: Balloon–payload structure, a flight train. Different components of the system are indicated in the figure.

### 2.3 Gas used for the balloon filling

Helium as an inert gas is considered the safest for filling the balloons, and we started our programme using helium, even for our tethered flight (Nayak et al., 2013). However, because it is expensive and due to scarcity of pure helium (Nuttall et al., 2012) in the world, we eventually switched over to a much cheaper commercial hydrogen. The price of a 10-$m^3$ helium cylinder (99.99% purity) is Rs. 17,000 as compared to Rs. 800 for a 7-$m^3$ commercial (99.99% purity) hydrogen cylinder. We buy gas (helium or hydrogen) for each launch from Sri Vinayaka Gas Agency, Bangalore, India.

### 2.4 Parachutes

Parachutes (Fig. 1) used in high-altitude ballooning are usually made out of *ripstop* nylon. We have obtained parachutes from Rocketman Parachutes Inc, USA, as they are of good quality and inexpensive, specially designed for high-altitude balloon payload recovery. We use parachutes of two different sizes, 7- and 8-foot diameter that have lift capacity of 4.0–4.9 kg and 5.4–6.8 kg, respectively. We also have small parachutes of 1-kg lift capacity from Aerocon Systems Inc, USA. In some flights we connect these parachutes serially.

## 2.5 Payload box

**First Model** The first payload model (Fig. 2, *Left*) was a rectangular structure made out of high-density Styrofoam with had dimensions of 20 × 20 × 20 cm. The base was 2 inch thick, and each side 1 inch thick. We wrapped the box in aluminum foil to provide the basic insulation, as well as the radar reflection. We, however, found that this shape gave rise to considerable spinning motion of the box.

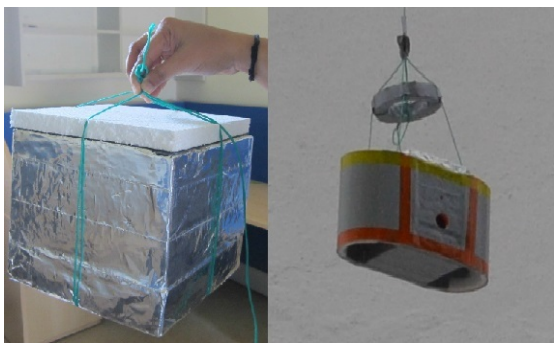

Figure 2: *Left*: First model; *Right*: Second model.

**Second Model** The second stage of development was to overcome the disadvantages of the first stage. The shape was selected to be cubic (15×15×15 cm) with wind-vane structure. The idea of having such a design is to align the payload always in one particular direction i.e. along the wind direction to limit the spinning of payload. This structure was made from lightweight plastic fibre and was open in the bottom and top to allow movement of the air through the vanes (Fig. 2, *Right*). This payload structure was much smaller and lighter than the first design. The thickness of the base and the sides was retained as in the previous model. However, the spinning motion was still substantial, and there was insufficient insulation for the instruments inside the box.

**Third Model** The third and final model (Fig. 2, *Right*) was the one that has been used for actual flights. To provide the insulation we have used Kapton tape, which withstands a temperature down to $-269°$C (Navick et al., 2004). The wind-vane structure with the same dimensions, box material and thickness as in the second model was retained. However, the top and bottom openings in the second model were covered with Styrofoam. A suitable ring structure was constructed to attach the payload to the parachute; this ring structure also helped in reducing the spinning of the payload. This payload was covered with bright coloured tapes to be easily noticeable during its recovery. We also pasted team members' mobile numbers on the payload box in case of it being found first by the public.

## 2.6 Flight Computer

For the initial flight, we used a BASIC STAMP 2 microcontroller (MC) module-based flight computer. It was found insufficient to support the necessary sensors and computations and we later switched to a Single Board Computer (SBC). For this purpose, the Raspberry Pi (RPi) SBC that reads and controls the sensors, calculates the system's attitude, and generates control commands, was developed in-house (Sreejith et al., 2014). It connects to a microelectromechanical system (MEMS) 9-axis IMU (MPU-9150, inertial measurement unit, InvenSense Inc, USA, http://www.invensense.com) containing two chips: the MPU-6050, comprising a 3-axis gyroscope, an accelerometer and a digital compass (Ak8975), and a Digital Motion Processor (DMP), which uses the proprietary InvenSense Digital motion fusion$^{TM}$. This IMU is commonly used in a variety of commercial applications including mobile phones, tablets and gaming platforms. The data stream from the MPU-9150 is fed into the I$^2$C (Inter-Integrated Circuit) port of the RPi. One of the great advantages of the RPi is the extensive library of open-source programs available for the platform. We use here the open source code *linux-mpu9150* (developed by the Pansenti software consultancy and development company, https://github.com/Pansenti/linux-mpu9150) to read the data from MPU-9150, com-bine the magnetometer data with the data from two other sensors, and calculate the attitude information in terms of Euler angles.

Celestial coordinates (RA and DEC) are calculated using the position data (latitude and longitude of the payload) obtained from the GPS sensor. We use a 20-channel GPS receiver from iWave[1] operating in the L1 frequency band (1575.42 MHz), which gives accuracy better than 10 meters on the ground and 20-30 meters in altitude. The entire system is powered by a 5V lithium polymer battery ($\sim$200 gms). We have written the $C$ code to combine the Euler angles with the GPS values and write the output in terms of RA and DEC to a file on a SD card. A photo of the sensor is shown in Fig. 3.

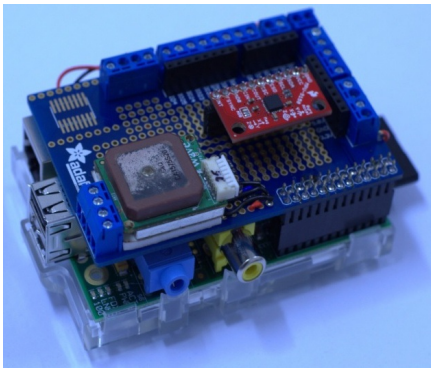

Figure 3: SBC with attitude sensor.

SBC interacts with different sensors and records data (temperature, attitude etc) and images at regular intervals throughout the flight on SD card. It also keeps a record of housekeeping files to verify the performance of the sensors onboard. The small size ($85.6 \times 56.5$ mm) and low power consumption (5 V, 1 A) of SBC makes it a suitable candidate for a flight computer.

## 2.7 Flight Termination Unit (FTU)

Balloons drift off if winds are high during the flight. To restrict the drift to another state or territory, we have designed two independent FTUs for the balloon system. One system is based on a timer circuit and placed below the parachute (Fig. 1) and the other – Arduino-based system with a GPS – is placed inside the main payload box. The FTU uses a thermal knife to cut the load line between the balloon and the parachute. We employ two types of the FTU systems. Details are described in the file In-HouseDevelopment.pdf available on our website: www.iiapballoongroup.wix.com/blue.

1. *Timer-based FTU*
   This mechanism consists of a timer circuit which is set for a duration of 2 to 3 hours depending on the type of experiment. It is initialized just before we release the balloon. The timer system is based on a tiny MC, which provides a trigger signal after the programmed time to heat up the nichrome wire wound across the load line to melt it, severing it for the deployment of the parachute.

2. *Geo-fencing based FTU*
   The geo-fencing system uses GPS to obtain latitude and longitude information. It sends an electrical signal to trigger the relay circuit which heats the nichrome wire to melt the load line, if the payload crosses the predefined latitude and longitude.

## 2.8 Payload

Depending on the scientific and technical objectives of every flight, the payloads differs. Some of the equipment is always present, such as camera, data sensors and tracking devices. Other equipment may be a telescope, spectrograph, pointing and stabilization platform, special sensors such as a Geiger counter, or an astrobiological set-up.

To capture the darkness of space, horizon and travel path of the balloon, we decided to use a camera onboard. Initially we used a Canon Ixus camera but found it to be too heavy for our purposes ($\sim$ 140 g, including batteries and memory card) and replaced it with a tiny RPi camera ($\sim$10 gms) programmed for a specific exposure time. We eventually found it useful to place two (or more) cameras onboard; one looking up monitoring the balloons, and other(s) looking either horizontally sideways or down, depending on the flight objectives. Additional sensors that we usually include in the payload are the environmental sensors: temperature, pressure, humidity.

---
[1]iWave Systems SiRF StarIII GSC3f GPS receiver, iWave Systems, India, http://www.iwavesystems.com

## 2.9 Flight Path Prediction

One–two days before the launch we calculate the balloon flight path using the Cambridge University Spaceflight Landing Predictor (http://predict.habhub.org/) together with our own codes written in MATLAB and IDL. The online predictor, as well as in-house developed codes, use data from the NOAA GFS[2] data and models for wind information. We calculate the atmospheric parameters from the standard model of atmosphere[3], which is an atmospheric model of how the pressure, temperature, density etc. vary over a wide range of altitudes. It specifies a set of base values, variables and equations, from which atmospheric parameters at different height from mean sea level can be calculated. The balloon information to use in the simulations is provided by the manufacturer. All our codes with explanations and how-to-run instructions are available on our website. Our launch site (CREST campus, IIA) coordinates are 13.1131 N, 77.8113 E, 960 m altitude.

# 3 Free-flying Balloon launches

## 3.1 Launch 1, (March 3, 2013)

This flight was our inception to the free-flying balloon experiments. To keep it simple, we used a small cubical shaped box to house our payload (Fig. 2, *Left*). The goal for the first flight was to develop the procedures for launching, tracking and recovering. We were also testing the electronic module and a GSM-based GPS tracker (SatGuide Tracker User Manual, SatNav Technologies, Hyderabad, India, 2012). It was an early morning flight scheduled for 6:30 am. We have used 3.4 m$^3$ of helium to fill the 1.2-kg balloon. The total weight of the payload was about 1 kg, so we have used 1-kg lift capacity parachute from Kohli Enterprises, Delhi (a free-sample parachute given for this launch).

---
[2]National Oceanic and Atmospheric Administration Global Forecast System, http://www.nco.ncep.noaa.gov/pmb/products/gfs/.
[3]ISA, https://en.wikipedia.org/wiki/International_Standard_Atmosphere

**Predictions for the flight**  To enable safe recovery of the payload, we carried out predictions and found the landing location to be near Malur, Karnataka (about 15 km from CREST campus). Fig. 4, plotted using Google Earth, shows the predicted path as well as launch, burst and landing locations.

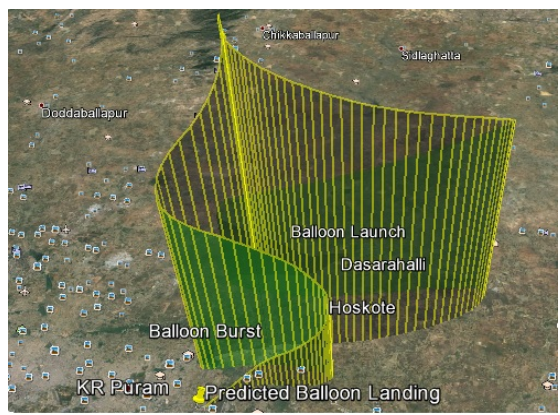

Figure 4: Prediction for Launch 1, March 3, 2013.

**Payload**  Our first low-cost flight computer, BASIC STAMP 2 MC module of Parallax Inc (Fig. 5, *Top*), was developed in-house to measure atmospheric parameters, such as pressure and temperature, along with latitude and longitude positions, using various sensors. These parameters could be recorded onto any USB device (e.g. pendrive). The bread-board model of the MC was used for the flight. The latitude and longitude positions could not be transmitted live because a radio license is required to carry out the live transmission over a radio band. To track the balloon, we obtained a small GPS unit called Satguide tracker (Fig. 5, *Bottom left*), which worked on GSM signals (Tata Docomo network). To capture images during the flight, we place a Canon Ixus 115HS camera horizontally on the payload box (Fig. 5, *Bottom right*).

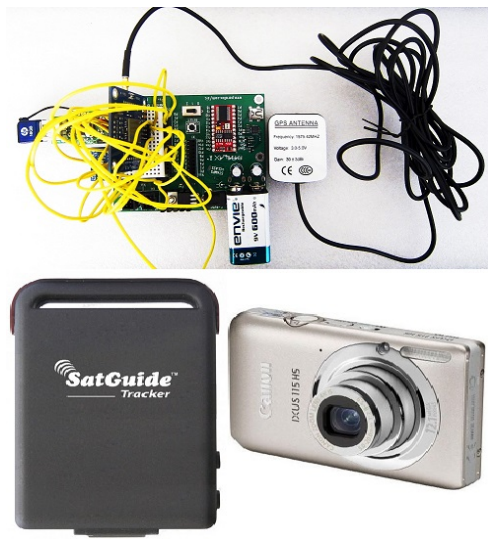

Figure 5: *Top*: Basic Stamp 2 MC board, *Bottom left*: Satguide tracker, *Bottom right*: Canon Ixus 115 HS camera used onboard.

**Recovery**  The FTU was programmed for 1 hour cut-off. We, however, lost the signal from the Satguide at 7.50 am, and the search at the predicted landing location failed though we were informed that the local residents saw the descending payload with the parachute. After two days, local residents called and informed us that they have found the payload (near the predicted location). It was recovered in a bad condition and we lost the parachute, camera and Satguide GSM tracker.

**Flight Analysis and Summary**  We lost the GSM signal due to poor network (Tata Docomo) in Malur-Hoskote area. We also noticed that GSM networks do not work at altitudes above 3 km. The FTU did work as the payload had descended successfully at the predicted location. Though we had recovered the MC, we did not recover any data. Since we did not find the performance of BASIC STAMP microcontroller to be satisfactory, we have decided to use SBCs for our future launches.

### 3.2   Launch 2, (June 30, 2013)

After a failure in tracking using only GSM trackers in the previous launch, we decided to add radio tracking in the second launch. We incorporated the radio tracker from Dhruva Space Pvt. Ltd.[4] to live-track the balloon (Fig. 1, bottom box). The total weight of the payload was 3.2 kg, therefore we have used two helium-filled balloons: 3-kg and 1.2-kg, using 3-kg as a master balloon (Fig. 6). Two 1-kg parachutes were connected serially. We included an in-house developed RPi-based attitude sensor to serve as the flight computer. The launch was carried out at 7:20 AM.

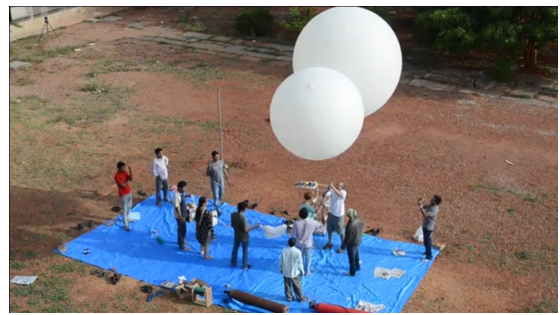

Figure 6: Launch 2, June 30, 2013, with two balloons.

**Predictions for the flight**  Using the available wind data for that day we found the predicted location of landing to be in Ramnagara district, Karnataka, about 120 km from CREST campus.

**Preparations for the launch**

  **Accuracy of the RPi sensor**  Testing of the in-house developed attitude sensor mounted on the telescope (Fig. 7) was carried out for this flight prior to the launch. For details of development and testing, see (Sreejith et al., 2014).

---

[4]Dhruva Space Ltd. http://www.dhruvaspace.org.

**Payload** We had two payload boxes this time.

- Our payload comprised the flight computer programmed to store data onboard. An attitude sensor with a webcam looking down was interfaced with the RPi. The attitude sensor continuously logged the RA and DEC of the payload pointing direction, as well as the latitude, longitude, UTC and altitude of the payload location. The webcam was taking low-resolution images every minute. A timer-based FTU, set for 90 minutes, was programmed to detach the payload from the balloon.

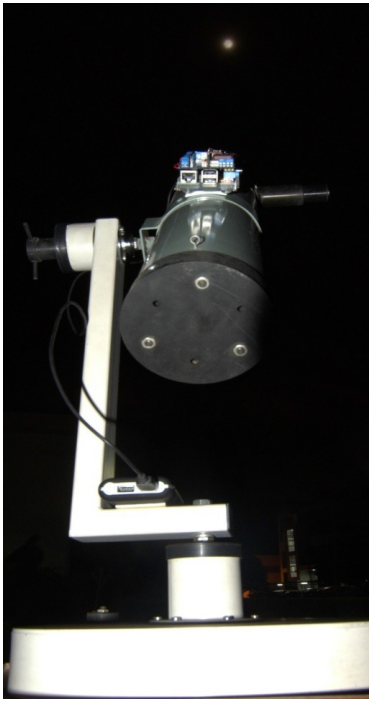

Figure 7: Attitude sensor mounted on the telescope for calibration.

- Dhruva Space payload box contained a telemetry unit and an RPi, interfaced with an IR camera. The plan was to compare the visible and IR images of the ground to find the chlorophyll content in Bangalore region. The telemetry unit, preconfigured with a call sign MT-AIO (Micro-Trak All-In-One) is a complete, self-contained, water resistant and portable 10-watt Automatic Packet Reporting System (APRS) tracker comprising a TinyTrak3 controller chip, a Byonics GPS2OEM GPS receiver, and a Standard Male antenna (SMA) (Fig. 8). It operates using 8 conventional AA batteries or a 12V power supply. The MT-AIO has been confirmed to run for nearly eight days when used with typical AA alkaline batteries, transmitting every two minutes.

**Telemetry Unit** The telemetry unit, preconfigured with a call sign MT-AIO (Micro-Trak All In One), was used in this launch for live tracking. It is a complete, self-contained, water resistant, portable, 10-watt Automatic Packet Reporting System (APRS) tracker, a TinyTrak3 controller chip, a Byonics GPS2OEM GPS receiver, and a Standard Male antenna (SMA), (Fig. 8). It operates using 8 conventional AA batteries or a 12V power supply. The MT-AIO has been confirmed to run for nearly eight days when used with typical AA alkaline batteries, transmitting every two minutes.

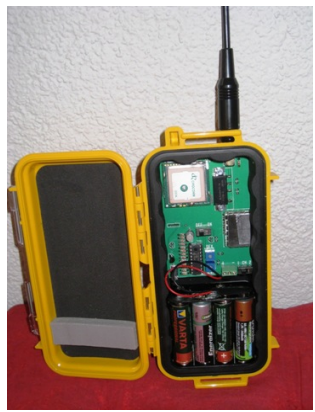

Figure 8: Dhruva Space telemetry unit in a *Pelican* case.

**Recovery** We lost the radio signal near Mugabala village, north of CREST, Hoskote; the last radio reading was at 11 km lateral distance. The payload was recovered after two weeks near a small village in Ramanagara district following a phone call from local residents, near the predicted landing location. We recovered all the electronic components, except the radio antenna, from the landing site.

**Flight Analysis and Summary** The objective of this flight was to test our attitude sensor and the radio communication link between the payload and ground station provided by the Dhruva Space. Radio signal was lost at 8:25 am, 1 hr after the launch, most probably due to the loss of the transmitting antenna. The last recorded altitude was 10 km (Fig. 9). However, the maximum altitude reached before the cut-off, as estimated from the ascent rate, was 22.3 km.

Our RPi failed 40 mins after the launch. The drop in temperature might have caused the power supply battery of the RPi to drain, probably because the electronic module (EM) was not properly insulated. We, however, have used the available attitude sensor data to plot the azimuth and elevation of the pointing direction (Fig. 10). It can be seen that the azimuth and elevation were varying randomly. We attribute these variations to the random nature of winds in the troposphere. Variation in the elevation was smaller which indicates that the swinging motion of the payload box was less pronounced. It is here, in the lowest part of the troposphere, called the planetary boundary layer (PBL), that the atmosphere experiences maximum turbulence and strong winds [Tong 2010].

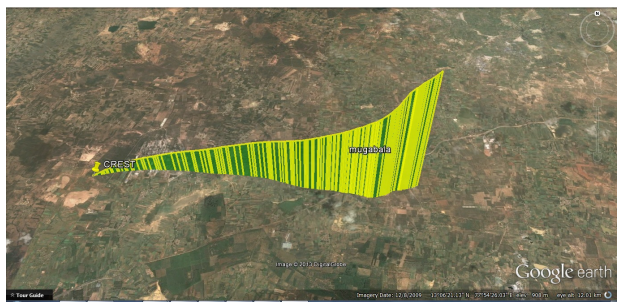

Figure 9: Launch 2, June 30, 2013, GPS plot of the flight path.

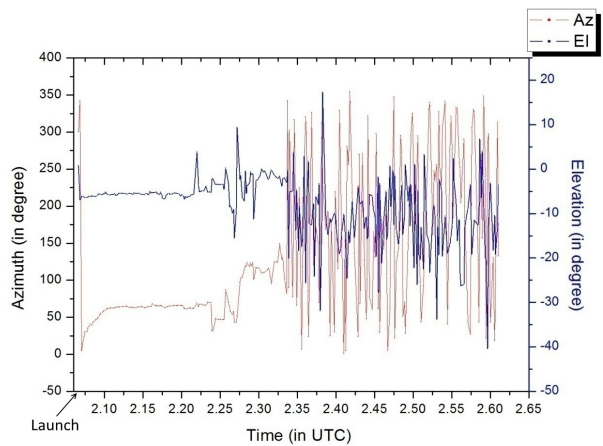

Figure 10: Azimuth-Elevation variations recorded on Launch 2, June 30, 2013. X-axis shows time of the flight in UT.

## 3.3 Launch 3, (September 3, 2013)

The September launch was a test flight in preparation to capture the images and UV spectra of the comet ISON, which was to have a closest approach to the Sun on November 28, 2013. The total weight of the payload was 2.4 kg, therefore we used one helium-filled 3-kg balloon and two 1-kg parachutes connected serially (Fig. 11). The launch was at 6.20 am and the payload landed at 7.40 am.

**Predictions for the flight** The predicted landing location was 25 km from CREST campus in the south direction, and the landing location was close to the predicted (Fig. 12).

**Preparation for the launch**

**Rope-ring mechanism** In the previous launches, the payload was subjected to jerks during the release. To ensure smooth release of flight train, we decided to use a rope-ring mechanism. This mechanism provided vertical suspension and stability of the payload at release. The rope connected to two poles was passed through the ring. The ring was used to support the whole balloon–payload structure. Balloons and the parachute were attached to this ring from opposite sides. This way, the payload was suspended vertically and stabilized before the launch. Finally, the rope was cut to allow the balloon to have a smooth launch.

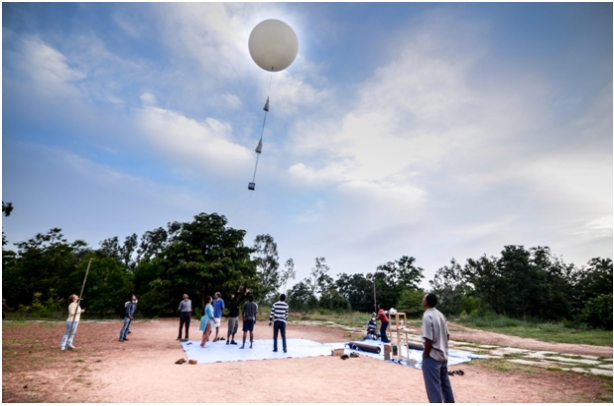

Figure 11: Launch 3, September 3, 2013, with the rope-ring mechanism.

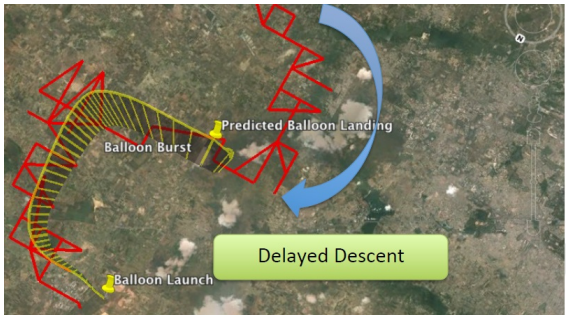

Figure 12: Prediction for Launch 3 on September 3, 2013.

**Payload** The payload consisted of one box containing a Medium Personal 1 (MP1) GSM tracker from Ingolabs, Hyderabad, India, with an Airtel SIM card, SBC as the flight computer, a temperature sensor and an FTU, set for 45 mins. Dhruva Space payload had a Radio AIO unit (same as in the previous launch) and an Arduino-based GPS-GSM tracker with a Vodafone SIM card, placed in the same box.

**Recovery** We lost the GSM signal when the balloon rose above 3 km. The radio unit was working all the time. The payload was recovered on the same day at 9.20 am in Oblapura village, Karnataka, 26 km away from the CREST campus. Both radio and GSM–GPS trackers transmitted the exact landing location of the payload. The complete video of this launch can be found on our website: www.iiapballoongroup.wix.com/blue.

**Flight Analysis and Summary** The predicted maximum altitude was 12 km. The altitude recorded by the radio tracker was 12.9 km (Fig. 13). The GPS positions, continuously transmitted by the radio tracker, were used to calculate the ascent and descent rate of the payload. The average ascent rate was found to be about 7 m/s, and the descent rate at touchdown was 5 m/s, as expected. Dhruva Space box recorded the drop in temperature inside the box which correlated with the measured drop in GPS transmitter voltage (Fig. 13, bottom row). The battery draining with decrease in temperature is the issue that we had to address in the following flights.

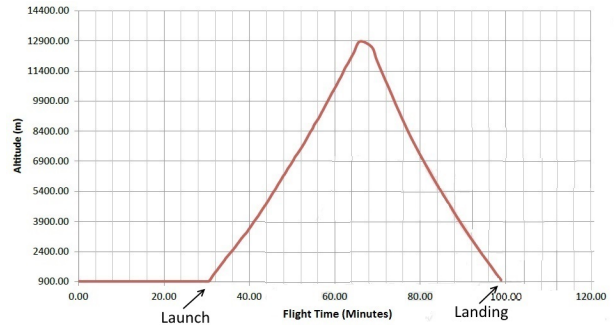

Figure 13: Data recorded on Launch 3, September 3, 2013. *Top left*: Change of altitude with time as recorded by the GPS onboard the AIO unit. *Top right*: Change in external temperature as recorded by the SBC temperature sensor. *Bottom*: Dhruva Space GPS transmitter inside temperature (*Left*) and voltage (*Right*) variation through the flight.

## 3.4 Launch 4, (October 13, 2013)

This launch was to test the gimbal (Model Aviation, Bangalore) — a 3-axis gyro-stabilization platform designed to point and stabilize the telescope onboard the flight (Fig. 14, *Top*). The total weight of the payload was about 5.5 kg, therefore we used two helium-filled 3-kg balloons and one 8-ft Rocketman parachute. The launch time was 6.15 am and the FTU was set for 90 mins cut-off.

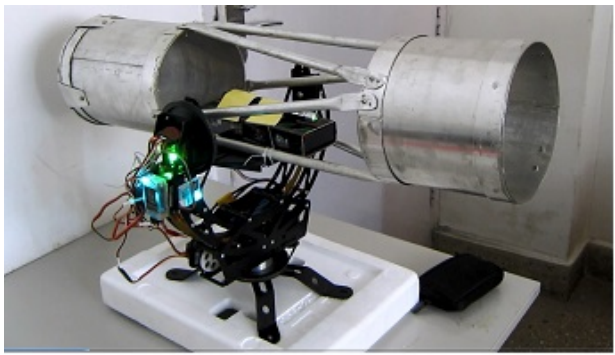

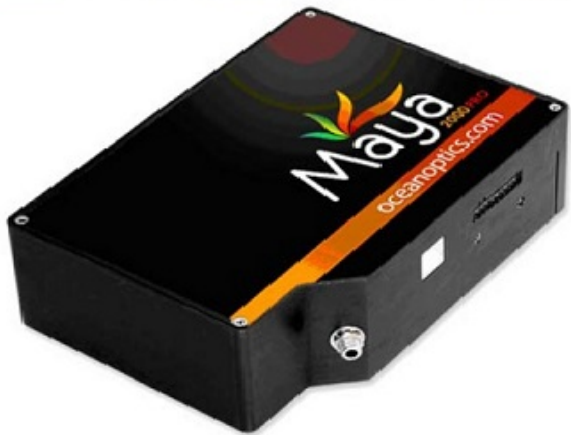

Figure 14: *Top*: Launch 4 Model Aviation gimbal with a designed in-house telescope mounted for testing. *Bottom*: UV Spectrograph (Mayapro 2000, Ocean Optics, USA).

**Predictions for the flight** The predictions for the flight path were carried out and the expected maximum altitude was 25 km. The balloon burst was expected to be close to Tiptur, Karnataka (Fig. 15).

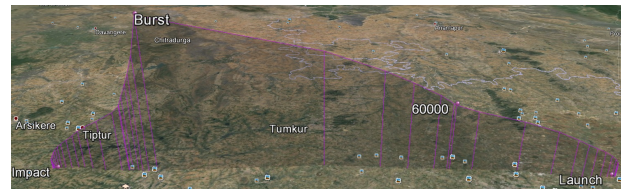

Figure 15: Predicted path of the balloon for Launch 4, October 13, 2013.

**Preparations for the launch** Laboratory tests where conducted to verify gimbal's stability of the pointing direction. The ground level tests were performed in the mechanical workshop by suspending the gimbal by a crane to provide conditions similar to the balloon flight. The azimuthal accuracy at a distance of about 13 meters was $\sim 13' - 15'$ and so was the accuracy in elevation at a distance of about 25.4 meters. The response time for a 45° angle slew was less than 1 second, and for angle of 90° or more, less than 2 seconds.

We also conducted insulation tests for the payload box prior to the launch by placing one open on top and one covered and taped Styrofoam boxes into a refrigerator with an ambient temperature of $-22\,°\mathrm{C}$ for 2 hours. An Arduino-based temperature data logger was kept in both boxes. Industry standard digital sensor with I$^2$C/SMBus interface LM75 was used to record the temperature inside the box every 15 seconds. The temperature plot is shown in Fig. 16.

We found that the internal temperature drops much faster in uninsulated box compared to the insulated box. Since the usual time of our flights is about 2 hours, we realized the need to properly insulate the payload boxes to ensure the working conditions for the electronics.

**Payload** We had two boxes this time. Our payload consisted of the gimbal (1.6 kg) and its battery (lithium polymer, $\sim$200 gms), RPi with attitude and two (internal and external) temperature sensors, two

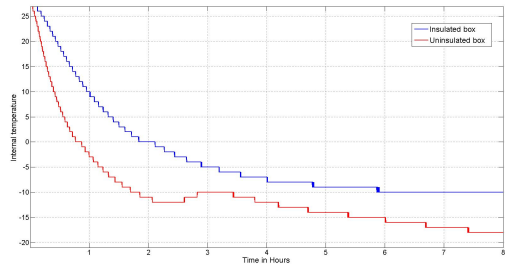

Figure 16: Temperature test results of the insulated (blue) and uninsulated (red) box.

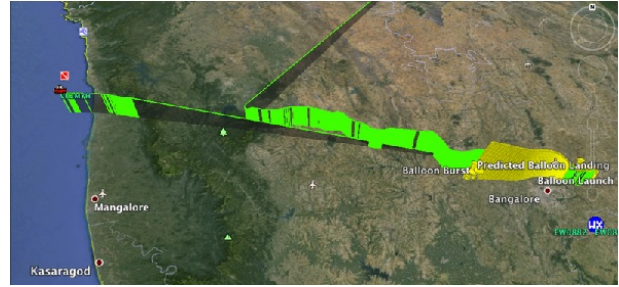

Figure 17: Launch 4, October 13, 2013. The actual path of the balloon.

GSM trackers (Airtel and Idea SIM card), Arduino-based timer FTU, a GoPro camera looking horizontally out and a webcam inside the box to monitor the gimbal. Dhruva Space payload box had the radio AIO unit.

**Recovery** The payload was found two weeks later by the fishermen near Udupi, Karnataka and returned to us. Most of the components, including gimbal and electronics, were lost. We were able to recover the attitude sensor data from the RPi, which we used in the calibration and development of pointing system (Sreejith et al., 2014).

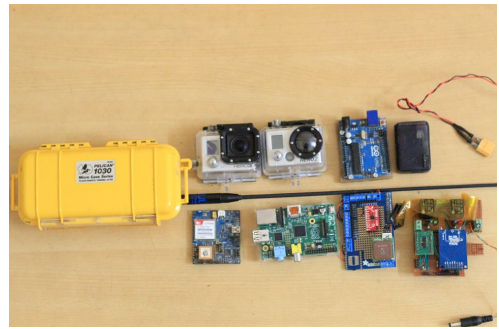

Figure 18: Launch 4, October 13, 2013. The components recovered from the sea.

**Flight analysis and Summary** The initial ascent rate was 7.5 m/s. From the recorded radio data we inferred that one balloon burst at the altitude of 20.5 km, while the remaining balloon descended to about 19–20 km altitude. The timer-based FTU failed, probably because the nichrome wire broke, and the balloon along with the payload eventually crossed into the Arabian sea via Karwar coast, Karnataka. We followed the balloon through the radio till the last reported position of about 150 km into the Arabian sea (Fig. 17). All recovered components are shown in Fig. 18.

### 3.5 Launch 5, (November 24, 2013)

The objective of this launch was to test the spectrograph (Fig. 14 *Bottom*) acquired to observe atmospheric lines and, in particular, the comet ISON at perihelion. The total weight was nearly 6 kg, therefore we used four helium filled balloons – one 3-kg, two 2-kg and one underfilled 1.2-kg balloon, and one 8-ft Rocketman parachute.

**Predictions for the flight** The wind data predictions gave the location of landing near Nandi Hills, Karnataka (Fig. 19).

**Preparations for the launch** We have used the new gimbal: a compact model with 2-axis stabilization, manufactured in National Institute of Technology (NIT), Surathkal, Karnataka, with a telescope mounted on it. The telescope body was fabricated by Cosmic Labs (Electronic city, Bangalore) based on our design. The telescope body was painted black to avoid scattering of the straylight. The light from the telescope was fed to the the spectrograph though the optical fibre placed in the prime focus (Fig. 20).

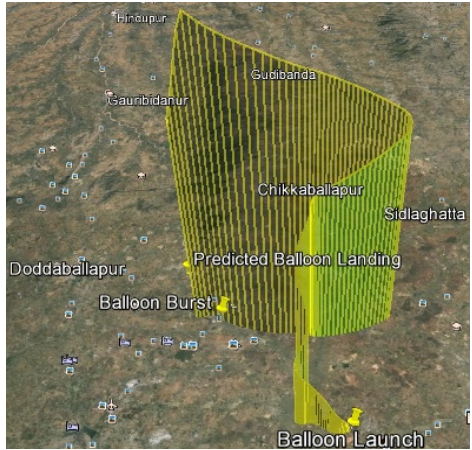

Figure 19: Prediction path for November 24, 2013, Launch 5.

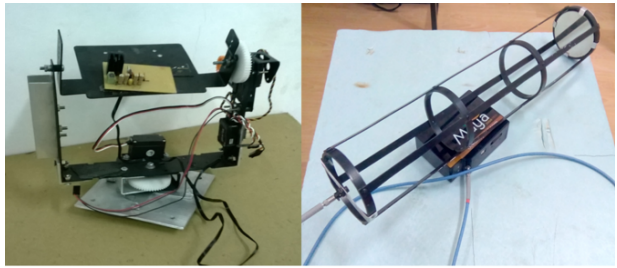

Figure 20: Launch 5, November 24, 2013. *Left*: Gimbal from NIT, Surathkal. *Right*: The telescope fabricated by Cosmic Labs (Electronic city, Bangalore) based on our design. The telescope body is painted black to avoid scattering of the straylight. The optical fibre collects light in the prime focus and feeds it into the spectrograph.

**Payload** The payload included three boxes. One box contained the spectrograph, telescope, gimbal, SBC with attitude and temperature sensors, two GSM trackers (Airtel and Idea) and a timer-based FTU, set for 90 mins. The second timer FTU, set for 120 mins, was placed in a separate box attached directly under the parachute to reduce the length of the wire. Dhruva Space payload box had the radio AIO unit.

**Recovery** The payload was recovered near Nandi hills, Bangalore, a little ahead of the predicted landing location. Three balloons burst before the programmed cut-off time and the remaining underfilled balloon served as a flag, which helped the recovery (Fig. 21). The payload was successfully tracked through radio and GSM trackers.

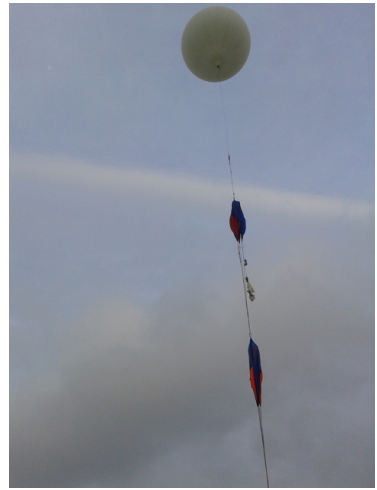

Figure 21: November 24, 2013: Recovery photo with the flag balloon and two parachutes.

**Flight analysis and summary** The launch was carried out at 4 am, and the payload was recovered by 7 am. The flight data showed that maximum altitude of $\sim 16$ km was achieved. All the electronics worked throughout the flight, however, the spectrograph did not record any data due to broken USB connection, most probably at launch.

## 3.6 Launch 6, (February 16, 2014)

This launch was carried out at 5 am to observe atmospheric lines at twilight. From this launch onwards we have switched over to hydrogen as lifting gas due to scarcity and high cost of helium. To provide the desired lift of 5.5 kg, we used three 2-kg hydrogen-filled balloons and one 8-ft parachute. Since the main objective of this flight was to study the atmospheric UV spectral lines, we decided to discard the heavy telescope structure and use the optical fibre to feed the light into the spectrograph (Fig. 22).

**Predictions for the flight**  As the high-altitude winds were predicted to be high on the launch day, we expected the balloon to drift faster. The predicted landing location was found to be 115 km away northeast from CREST campus.

**Payload**  The main payload was the spectrograph with the optical fiber (serving as an aperture) placed in the box horizontally to point at the horizon. The USB camera was placed on top of the payload box to monitor the balloons and an RPi camera was placed horizontally to scan at the horizon. The upward camera was programmed to capture images of the balloons every 30 seconds, and RPi camera was programmed for every 10 seconds. We had two FTUs: one timer-based and one geo-fencing. Since our scientific objective was observation of atmospheric lines, we did not require a stabilization platform (gimbal) for this flight.

**Recovery**  The balloons burst 90 mins after the launch and the payload landed around 130 km from CREST campus along the predicted direction. The payload was recovered successfully with all the electronics in working condition.

**Flight analysis and summary**  The maximum altitude obtained was 26.916 km. The upward USB camera recorded the premature (before reaching the burst altitude) balloon burst (Fig. 23, *Top left*). We also obtained the photographs from the horizontally placed RPi camera throughout the flight. In Fig. 24 we show images of the Moon, sunrise and Earth's horizon with stratospheric clouds from the RPi camera.

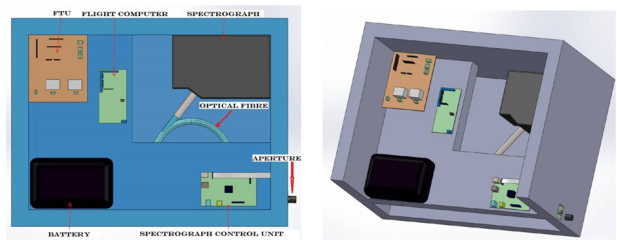

Figure 22: Launch 6, February 16, 2014. *SolidWorks* design of a payload box with the spectrograph and optical fibre. *Left*: Top view. *Right*: 3-D view.

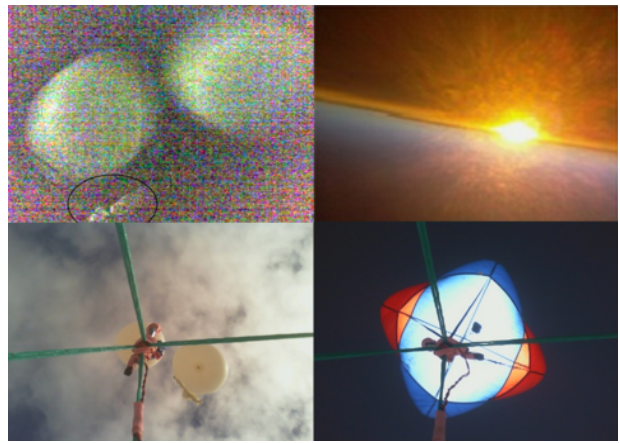

Figure 23: Launch 6, February 16, 2014. photos from upward USB camera. *Top left*: observed balloon burst (encircled). *Top right*: Sunrise. *Bottom left*: remaining balloons. *Bottom right*: open parachute during descent.

## 3.7 Launch 7, (May 4, 2014)

This launch took place at 9.37 am with the same payload as in the previous flight using hydrogen-filled three 2-kg balloons and one 1.2-kg underfilled balloon to reproduce the conditions of the Launch 5 (Fig. 25), where one underfilled balloon aided in the recovery of the payload. The total weight of the payload was 5 kg.

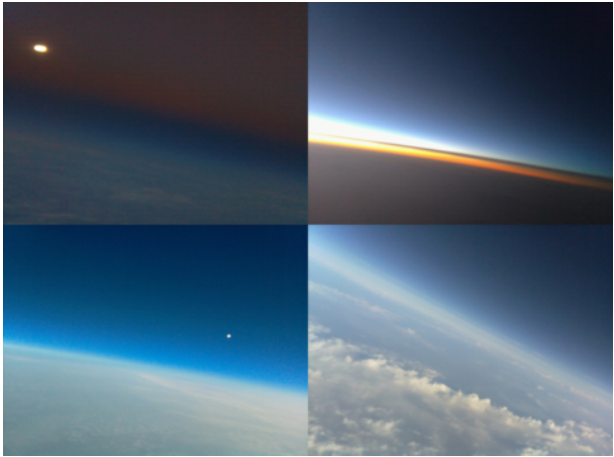

Figure 24: Launch 6, February 16, 2014. Photos taken by the horizontal RPi camera: *Top left*: the Moon, *Top right*: the sunrise, *Bottom*: Earth horizon at 25 km.

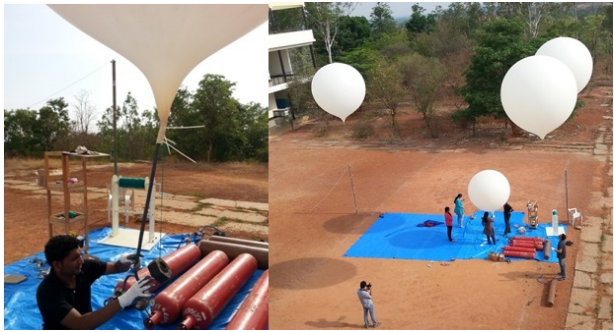

Figure 25: Launch 7, May 4, 2014. *Left*: Measuring balloon lift. *Right*: 4 balloons used for launch at 9.15 am.

**Predictions for the flight** The predictions showed the balloon path along Bangalore-Mangalore highway about 120 km from CREST campus.

**Preparations for the flight**

- *Spindle Fabrication* A rope spindle was fabricated in IIA workshop from mild steel. It served better purpose for the rope-ring mechanism. The spindle can be seen in Fig. 25, *Left*.

- *Temperature tests carried on April 16, 2014*

  Worried about the several failures of the FTUs, we have decided to place the FTU in the specialized *Pelican* case[5]. These are the airtight and watertight plastic containers that also include a barometric relief valve to sustain internal pressure during the environment changes. We have conducted the temperature tests on the *Pelican* case at very low temperatures. Two *Pelican* cases, one with the cut-down circuit (Fig. 26) and the other with the temperature sensor (MSR145 temperature data logger[6]), were kept for 1 hour 20 mins in a low-temperature chamber at the MRDG Dept, IISc, Bangalore, at an ambient temperature of $-80\,°\mathrm{C}$ to simulate the conditions at high altitudes. At the end of the experiment, the cut-down circuit was still working (Fig. 27). The plot shows that the temperature inside the box had reached around $-60°$ C. Although the datasheet for the FTU circuit put the working temperature on the FTU at above $-40°$ C, it was still working at $-60°$ C.

**Payload** The payload remained the same as in the previous flight with the main objective to observe atmospheric lines and airglow.

---

[5]Trademark of Pelican Products, Inc., http://www.pelicancases.com/Micro-Cases-C1.aspx

[6]The miniaturized universal Data Logger (MSR145) consists of the pressure sensor, temperature sensor, humidity sensor and 3-axis accelerometer (X, Y and Z axes) for continuous sensing and storage of atmospheric parameters such as pressure, temperature and relative humidity (RH). The Data Logger can be programmed to record the readings at any time interval. The compactness (20 × 15 mm), light weight (∼16 gms) and durable design of the MSR145 makes it ideally suited for routine monitoring of the environment the balloon passes through. The Data Logger is equipped with the software to plot and calculate the atmospheric data (MSR145 User Manual, MSR Electronics GmbH, Switzerland, 2008 July, Version 4.00).

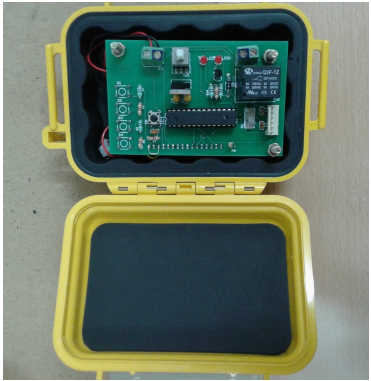

Figure 26: *Pelican* case with the FTU for the temperature test on April 16, 2014. The FTU battery is wrapped with Kapton tape and placed below the circuit board.

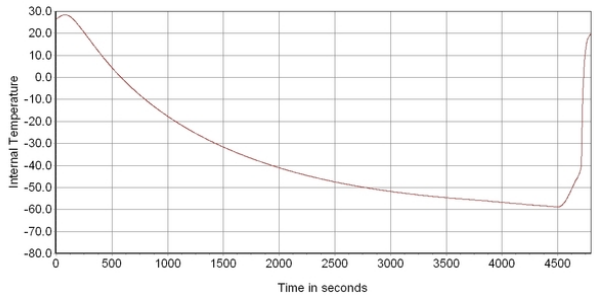

Figure 27: Temperature test result carried out on April 16, 2014.

**Recovery** The payload landed in Kadehalli area near the Bangalore–Mangalore highway. The local residents gave a call to the team and we recovered the payload next day with all the equipment intact.

**Flight analysis and Summary** All data from the payload was recovered and analyzed. From the upward looking USB camera photographs we inferred that one balloon burst 15 mins after the launch, its rope got entangled with the parachute and prevented it from opening. The FTU worked but since the ropes got entangled, no separation occurred. Two other balloons burst at the predicted time. The remaining balloon was diffusing the gas and acted as a parachute for the descend of the payload. The spectrograph recorded the scattered UV solar spectrum, the attitude and temperature information was recorded by the SBC. The maximum altitude achieved was 25 km. We used the spectrograph data to study the variation of the scattered UV solar flux with altitude and for the detection of the variation in the strength of the airglow lines.

### 3.8 Launch 8, (June 15, 2014)

The experiment was launched at 8.43 am with the same payload as in previous flights. Two 2-kg balloons were filled with hydrogen to lift the total weight of 4.5 kg.

**Predictions for the flight** This time the predictions showed the same direction as on May 4th, along the Bangalore–Mangalore highway, as wind patterns are generally the same for this particular time period (Fig. 28, *Top*).

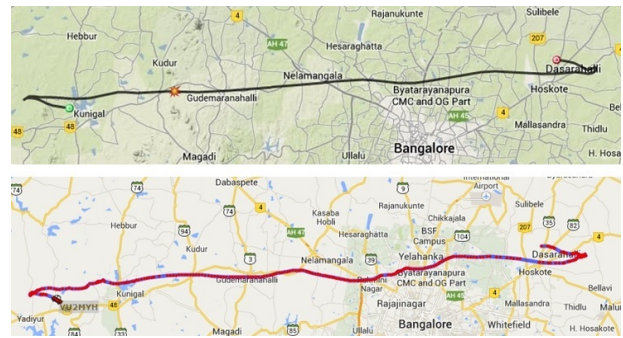

Figure 28: Launch 8, June 15, 2014. *Top*: Predicted path of the balloon travel. *Bottom*: Actual path travelled by the balloon.

**Preparations for the flight**

**Temperature test carried on May 28, 2014** We have repeated the temperature test of April 16th to check the survival of FTU circuits at low temperatures for extended time. The FTU circuit was programmed to trigger at 5 different times: 1, 1.5, 2, 2.5 and 3 hrs. The FTU circuit worked every time, and the battery voltages were recorded (Table 1).

Table 1: Cut-down time and battery voltage data.

| Cut-down Time(Hrs) | Battery Voltage (V) |
| --- | --- |
| 1 | 7.61 |
| 1.5 | 7.38 |
| 2 | 7.14 |
| 2.5 | 6.95 |
| 3 | 6.99 |

This shows that the FTU battery lasts at least 5 cut-downs and the circuit can keep operating inside a *Pelican* case at down to $-80\,°C$ outside temperature.

**Payload** The payload included the UV spectrograph, two RPi-s, attitude and temperature (inside and outside) sensors, USB camera facing up towards the balloons, RPi camera placed horizontally, 3 GSM trackers (with Airtel, Idea and Vodafone SIM cards), 2 FTUs in separate boxes (one set for 2 hrs and other for 2.15 hrs) and Dhruva Space radio tracker.

**Recovery** The payload landed in Hulipura village near the Bangalore–Mangalore highway and was collected by local residents before the recovery team reached the spot. The residents have returned the payload to us in perfect condition.

**Flight Analysis and Summary** The launch was carried out smoothly with all the systems ready on time. The actual path was exactly as the predicted path (Fig. 28). One balloon burst prematurely within 10 mins of the launch. Second balloon burst at 10.20 am. The first FTU worked at set time of 10.32 am. The RPi-1, which included the spectrograph control software and RPiCam, and the RPi-2, which included the USB camera, attitude and temperature sensors, worked continuously throughout the flight. The temperature plot for the internal sensor showed the maximum temperature of $+53\,°C$ (Fig. 29), within the working range of the electronics. The maximum altitude reached was 28.48 km. The flight and science data from this launch is being used in our studies of the airglow, atmospheric lines and the dependence of the flight parameters at launch on the balloons performance. In Fig. 30 we show the example of the variation of scattered solar flux with altitude. This is part of the work in calculation of the variation of UVA (400–315 nm) and UVB (315–280 nm) scattered solar flux with height in troposphere and in stratosphere and the line strength analysis.

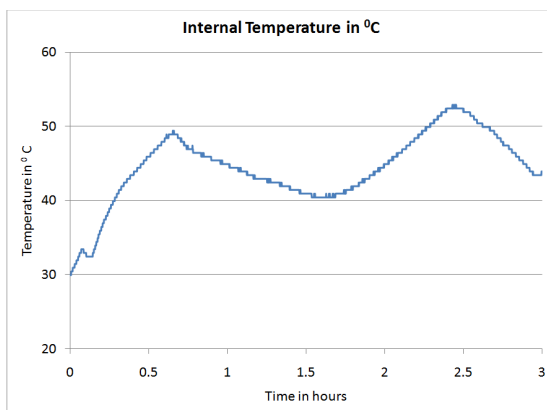

Figure 29: Payload box inside temperature recorded on Launch 8, June 15, 2014.

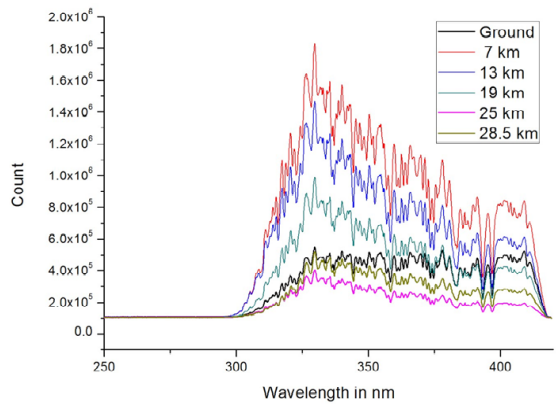

Figure 30: Launch 8, June 15, 2014. Variation in atmospheric spectra with increase in altitude.

## 3.9 Launch 9, (October 12, 2014)

The launch was scheduled for 6.30 am. Three 2-kg balloons were used to lift the total payload weight of 5 kg (Fig. 31).

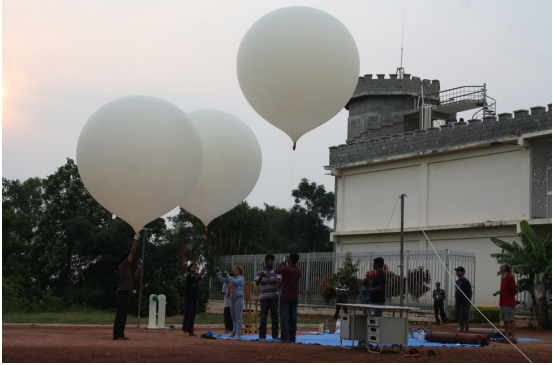

Figure 31: October 12, 2014, Launch 9 at 6.15 am.

**Predictions for the launch**  The predicted landing location was 110 km away west from CREST campus, near Savandurga forest, Karnataka.

**Preparations for the launch**  Prior to the launch, extensive vibration/shock tests for the complete payload were performed in the lab. In addition, the usual temperature tests were carried out, and the GSM trackers were tested for the performance.

**Payload**  The payload remained the same as in the previous launch: UV spectrograph, RPi camera, USB camera, attitude sensor, two temperature sensors, one inside and and one outside the payload box, two GSM trackers (Airtel and Aircel SIM cards), and the FTUs in two separate boxes, one set for 2 hrs and another for 2.15 hrs cut-off time. In addition, for redundancy, we have placed an android phone (with Idea SIM card) into the payload.

**Recovery**  The payload landed almost 110 km away from CREST campus at predicted location, near Savandurga forest, Karnataka. Though the radio tracker stopped working 10 mins after of the launch, the Airtel GSM tracker worked throughout the landing and transmitted the exact location of the payload, which was successfully recovered. The android phone also provided the location but with low accuracy.

**Flight Analysis and Summary**  One balloon burst prematurely within 20 mins of the launch. The other two balloons burst 90 mins after the launch. The radio tracker failed to work in this launch. GSM trackers were used to locate and recover the payload. All the scientific instruments on the payload worked throughout the flight and the data was successfully recovered. The scattered solar spectrum at different times was obtained during this flight. The maximum achieved altitude was 26 km as estimated from simulations and temperature readings.

# 4 Ongoing and Future work

We are currently analyzing the scattered solar UV spectra obtained from three launches (Launch 7–9). One of our objectives is to detect the airglow lines and to establish the altitude dependence of the strength of the lines. The collected flight data is being used to model the performance of the flights with multiple balloons, the dependence of the ascent/descent rates on the number of balloons, gas volume, payload weight, etc.

To reduce the payload weight, we have acquired Replicator2 Desktop 3-D printer (MakerBot, http://www.makerbot.com) using a biodegradable polymer polylactic acid (PLA) to print mechanical structures, such as containers, platforms, etc. We have developed and printed the 2-axis stabilization and pointing platform onto which the optical fibre serving as an aperture is mounted to observe the extended astronomical sources (Fig. 32). We are currently testing and calibrating this platform (Nirmal et al., 2015) for which we have developed the attitude sensor (Sreejith et al., 2014). For additional positional precision, the ballooning/minisatellite community is using sun sensors. Since one of our objective is observation of astronomical sources, we need to observe in the night, therefore we require a star sensor. Commercially available star sensors are very expensive, at 50,000 to 100,000 USD. Our current work is on the development of the lightweight low-cost star camera-cum-sensor with image-intensified CMOS sensors and associated optics to provide accurate pointing and stability for the payload (Sarpotdar et al., 2014).

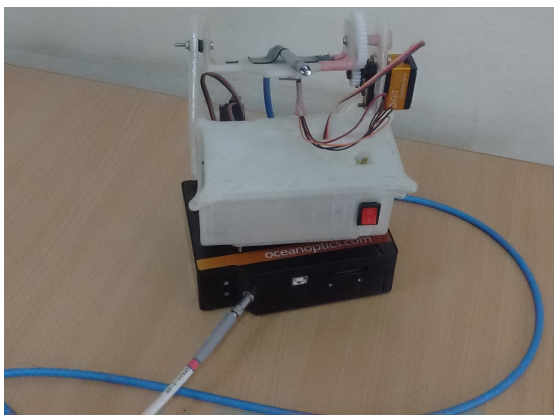

Figure 32: 3-D printed 2-axis stabilization and pointing platform for observations of airglow and atmospheric lines. The light from the optical fibre serving as an aperture is fed into the spectrograph.

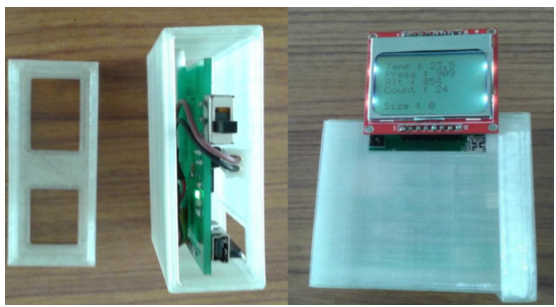

Figure 33: 3-D printed box for the data logger. *Left*: Data Logger box with a cover (left). *Right*: data logger circuit inside the printed box, with a removable LCD display.

The Mayapro 2000 spectrograph from Ocean Optics that we were using for our airglow and atmospheric lines observations has nevertheless a low quantum efficiency in the UV, being a CCD-based system. In addition, its maximum integration time is 5 seconds and a wavelength resolution is more than 1 nm. Therefore, we currently developing a new near-UV spectrograph, using off-the-shelf components, for the dedicated UV studies. The new spectrograph will have an image-intensified CMOS detector, thus enhancing UV sensitivity. The MCP-based detectors also provide a low-noise detector system. It will have a factor of two improvement in wavelength resolution (around 0.5–0.6 nm) compared to the existing one, along with more than 8 times improvement in the integration time. The system will be adjustable in the sense that it will be possible to interface it with different telescope configurations through either an optical fiber or a slit of appropriate dimensions.

Since the MSR145 datalogger is every expensive (∼Rs. 90,000), we have developed a data logger in-house and printed its box on a 3-D printer Fig. 33. The design was performed with SolidWorks 3-D design software (http://www.solidworks.in). Details are described in the file In-HouseDevelopment.pdf available on our website: www.iiapballoongroup.wix.com/blue.

The future work goals include:

- Development of two-way communication link between the payload and the ground for the live transmission of scientific data.

- Development of the image-intensified UV detector on telescopic system for spectroscopic and imaging applications to fly onboard the balloon payload (Ambily et al., 2015; Sreejith et al., 2015).

- Design and development of the astrobiological set-up for collecting the stratospheric air and dust samples.

# 5 Conclusions

We present the overview and some results of our HAB programme experiments carried out in IIA CREST campus, Hoskote. In spite of many difficulties in running this programme due to a tight weight and cost budget, we show that it is still possible to achieve scientific results. The instruments used in these experiments usually give satisfactory results, and we continue our parallel programme of instrument development. We use the GPS trackers along with radio and generally follow the balloon path for payloads retrievals. Our experience show that our predictions for the balloon/payload path are generally correct and the payloads are always found close to the predicted locations.

# Acknowledgments:

The authors are thankful to Air Force stations (Mekhri Circle, Yelahanka, Chimney Hills, Bangalore), HAL, Chennai Airport Authorities and Jakkur Aerodrome for providing the necessary NOC to carry out our free-flying balloon experiments. Authors thank Dr. Annapoorni Rangarajan and students of her laboratory, MRDG, IISc, for providing low-temperature chamber facilities for temperature tests and a UV-lamp for spectrograph calibration.